\documentclass[prd,nofootinbib,preprint,floatfix]{revtex4}

\usepackage{amsmath,amssymb}
\usepackage[normalem]{ulem}
\usepackage{graphicx}
\usepackage{array}
\usepackage{color}

\newcolumntype{C}{>{~$}c<{$~}}
\newcolumntype{R}{>{~$}r<{$~}}

\definecolor{red}{cmyk}{0,1,1,0.4}

\preprint{\vbox{%
\hbox{\bf YITP-SB-05-06}}}
\begin{document}
\vspace*{.25in}
\title{Mass Varying Neutrinos in the Sun}
\author{Marco Cirelli}
\email{marco.cirelli@yale.edu}
\affiliation{
Physics Dept. - Yale University, New Haven, CT 06520, USA}
\author{M.C.~Gonzalez-Garcia}
\email{concha@insti.physics.sunysb.edu}
\affiliation{
C.N.~Yang Institute for Theoretical Physics,
SUNY at Stony Brook, Stony Brook, NY 11794-3840, USA
\\
IFIC, Universitat de Val\`encia - C.S.I.C., Apt 22085, 
E-46071 Val\`encia, Spain}
\author{Carlos Pe\~na-Garay}
\email{penya@ias.edu}
\affiliation{
School of Natural Sciences, Institute for Advanced Study,
Princeton, NJ 08540, USA
\vspace*{.25in}}

\begin{abstract}
In this work we study the phenomenological consequences of the
dependence of mass varying neutrinos on the neutrino density in the Sun, 
which we precisely compute in each point along the neutrino trajectory.
We find that a generic characteristic of these scenarios 
is that they establish a
connection between the effective $\Delta m^2$ in the Sun and the
absolute neutrino mass scale. This does not lead to any new 
allowed region in the oscillation parameter space. On the contrary, 
due to this effect, the
description of solar neutrino data worsens for large absolute mass.  
As a consequence a lower bound on the level of degeneracy
can be derived from the combined analysis of the solar and KamLAND data. 
In particular  this implies that the analysis favours normal over inverted 
mass orderings. These results, in combination with a positive independent 
determination of the  absolute neutrino mass, can be used
as a test of these scenarios together with a precise determination 
of the energy dependence of  the survival probability  of solar neutrinos, 
in particular for low energies. 
\end{abstract}
\maketitle
\section{Introduction}
\label{sec:intro}

Ref.~\cite{dark1} recently discussed the possibility that mass varying
neutrinos (MaVaNs) can behave as a negative pressure fluid which
contributes to the origin of the cosmic acceleration.  In
particular the authors consider a scenario in which the neutrino mass
arises from the interaction with a scalar field, the acceleron ${\cal
A}$, whose effective potential changes as a function of the neutrino
density. This establishes a very intriguing connection between
two recent pieces of evidence for New Physics --the indirect
observation of Dark Energy and the confirmation of neutrino masses and
oscillations-- that are both suggestively characterized by a similar
mass scale.
Besides the possible interesting cosmological
effects~\cite{dark1,cosmo consequences},
from the point of view of neutrino oscillation phenomenology the
unavoidable consequence of this scenario is that the neutrino mass
depends on the local neutrino density and therefore can be different
in media with high neutrino densities such as the Sun.

A subsequent work Ref.~\cite{dark2} also investigated the possibility
that neutrino masses depend on the visible matter 
density as well. Such a dependence would be induced by  
non-renormalizable operators which would couple the acceleron also to the 
visible matter and could lead to interesting phenomenological consequences
for neutrino oscillations~\cite{dark2,Zurek,Barger}. However,  
unlike the dependence on the local neutrino density, 
which is an unavoidable consequence of the proposed 
MaVaNs mechanism,  the possible dependence on the visible matter density 
is strongly model-dependent.
In principle it could be vanishingly small since  so far the only information 
on the effective acceleron-matter couplings are upper bounds from 
tests on the gravitational inverse square law. 

Consequently, in this work we concentrate on the phenomenological 
consequences associated to the unavoidable dependence of MaVaNs 
on the neutrino density in the Sun. 
We find that a generic feature of these scenarios is that they establish a
connection between the effective $\Delta m^2$ in the Sun and the
absolute neutrino mass scale $m_{01}$. Due to this effect, the
description of solar neutrino data worsens for large $m_{01}$.
In other words,
a lower bound on the level of degeneracy
$\Delta m^2_{21,0}/m^2_{01}$ can be derived from the
combined analysis of the solar~\cite{chlorine,sagegno,gallex,sk,sno}
and KamLAND data~\cite{kamland}. For the realization
considered in this work, the 3$\sigma$ bound is 
$\Delta m^2_{21,0}/m^2_{01}> 1 $ 
from the analysis of solar plus KamLAND data. 
In particular this implies that these scenarios 
favour normal mass orderings as for  inverse mass orderings   
$m_{01}^2\simeq \Delta m^2_{\rm ATM} \gtrsim 10^{-3}$ eV$^2$ which 
already implies
$\Delta m^2_{21,0}/m^2_{01}\lesssim 0.1$. Conversely, 
the constraint on $m_{01}$ will allow a test of the validity of 
these scenarios in the event of a positive determination of the 
absolute neutrino mass scale from independent means.

The outline of the paper is as follows. In Sec.~\ref{sec:mass} we
evaluate the density profile of neutrinos in the Sun in the SSM and
discuss the results on the expected size of the neutrino mass shift
induced for different forms of the scalar potential.  Section
~\ref{sec:osc} contains our results for the effective neutrino mass
splitting in the Sun and the modification of the solar neutrino
survival probability. Finally in Sec.~\ref{sec:analysis} we illustrate
the generic quantitative consequences of these scenarios by presenting
the results of an analysis of solar (plus KamLAND) data for a
particular realization.

\section{Mass Varying Solar Neutrinos}
\label{sec:mass}

For most purposes in this section, the derivation of the effective
neutrino mass in the presence of the solar neutrino background  
can be made in a model independent way  using the neutrino mass 
$m_\nu$ as the dynamical field (without making explicit use of the 
dependence of $m_\nu$ on  the acceleron field ${\cal A}$). 

In this approach  at low energies the effective 
Lagrangian for $m_\nu$ is   
\begin{equation} 
\mathcal{L}= m_\nu \bar\nu^c
\nu + 
V_{tot}(m_\nu) \, ,
\end{equation}
where 
$V_{tot}(m_\nu)=V_\nu(m_\nu)+V_0(m_\nu)$ contains the contribution to the 
energy density both from the neutrinos as well as from the scalar potential. 
The condition of minimization of $V_{tot}$ determines the physical 
neutrino mass.

The contribution of a neutrino background to the energy density is given by 
\begin{equation} 
V_\nu = 
\int
\frac{d^3 k}{(2\pi)^3} {\sqrt{k^2 + m_\nu^2}}\ f(k)\ ,
\label{eq:deltav}
\end{equation} 
where $f(k)$ is the sum of the neutrino and antineutrino
occupation numbers for momentum $k$. $V_\nu$ receives contribution 
from the cosmological Big Bang remnant neutrinos as well as from 
any other neutrinos that might be present in the medium. Thus in general
\begin{equation}
\label{Vnu}
V_{\nu}(m_\nu) = 
V_{C\nu B} + V_{\nu, \rm medium}
= 
m_\nu \ n^{C\nu B} + V_{\nu, \rm medium}\, ,
\end{equation}
where we have used that in the present epoch relic neutrinos are 
non relativistic. $n^{C\nu B}=112$ cm$^{-3}$  for each neutrino species.
In 
a medium like the Sun, which
contains an additional background of  relativistic neutrinos, 
$V_{\nu,\rm medium}$  is given by Eq.~(\ref{eq:deltav}).
Notice that in writing Eq.~(\ref{eq:deltav}) we have neglected 
the possible dependance of the neutrino mass on the ordinary matter 
density, mediated by the acceleron field~\cite{dark2}. In the language 
of~\cite{dark2}, this implies that we are assuming that $\lambda_{B} \ll 
10^{-3}$, where  $\lambda_{B}$ is the coupling of the scalar field with 
baryonic matter.

Thus in the Sun, the condition of minimum of the effective potential 
reads
\begin{equation}
\frac{\partial V_{tot}(m_\nu)}{\partial m_\nu}\rfloor_{m_\nu} = 0
\quad \Rightarrow 
\quad  V'_0(m_\nu) +n^{C\nu B}(1+ m_\nu \,A)=0 \, ,
\label{eq:minimization}
\end{equation}
where we have defined the average inverse energy parameter normalized
to the CMB neutrino density
\begin{equation}
A\equiv 
\frac{1} {n^{C\nu B}} 
\int \frac{d^3 k}{(2\pi)^3} \frac{1}{\sqrt{k^2 + m_\nu^2}} \,
f_{\rm Sun}(k)\, .
\label{eq:a}
\end{equation} 

In the SSM the 
distribution of relativistic electron neutrino sources 
in the Sun is assumed to be spherically symmetric and it
is described in terms of radial distributions $p_i(r)$  
for $i=pp$, $^7$Be, $N$, $O$, $pep$, $F$, and $^8$B fluxes.  
As a consequence, the density of neutrinos in the Sun 
is only a function of the distance from the center of the Sun, $x$. 
It is computed integrating over the contributions at point $x$ due to the
neutrinos isotropically emitted by each point source, as:
\begin{equation}
n^{\rm Sun} (x) ~=~ \sum_i K_i \frac{2\pi}{x} \int dr ~ r ~
\log \frac{x+r}{|x-r|}p_i(r)~.
\label{eq:density2}
\end{equation}
$K_i$  are constants determined by normalization  
to the observed neutrino fluxes at the location of the Earth as:  
\begin{eqnarray}
n^{\rm Sun} (x) ~&=&
~ \sum_i 
\frac{(1 AU)^2}{2 R_\odot ^2} 
\frac{1}{x}~\frac{\Phi_{\nu,i}}{c} \int d r ~ 4\pi 
 r ~\log \frac{x+ r}{|x- r|}   p_i( r)\nonumber\\
&=&
4.6 \times 10^4\, {\rm cm^{-3}}\,
\frac{1}{x} \sum_i \alpha_i\int d r ~ 4\pi 
 r ~\log \frac{x+ r}{|x- r|}   p_i( r)\, .
\label{eq:density7}
\end{eqnarray}
Both $r$ and $z$ are given in units of $R_\odot$ 
so $\int_0^1 4\pi r^2 p_i(r)=1$ and $\alpha_i=\Phi_{\nu,i}/\Phi_{\nu,pp}$.
We use in our calculations the fluxes from 
Bahcall, Serenelli and Basu 2005
BS05(OP)~\cite{BS05}, 
 and the corresponding production point distributions
$p_i(r)$~\cite{jnbwebpage}.
 
\begin{figure}[ht]
\includegraphics[width=3.2in]{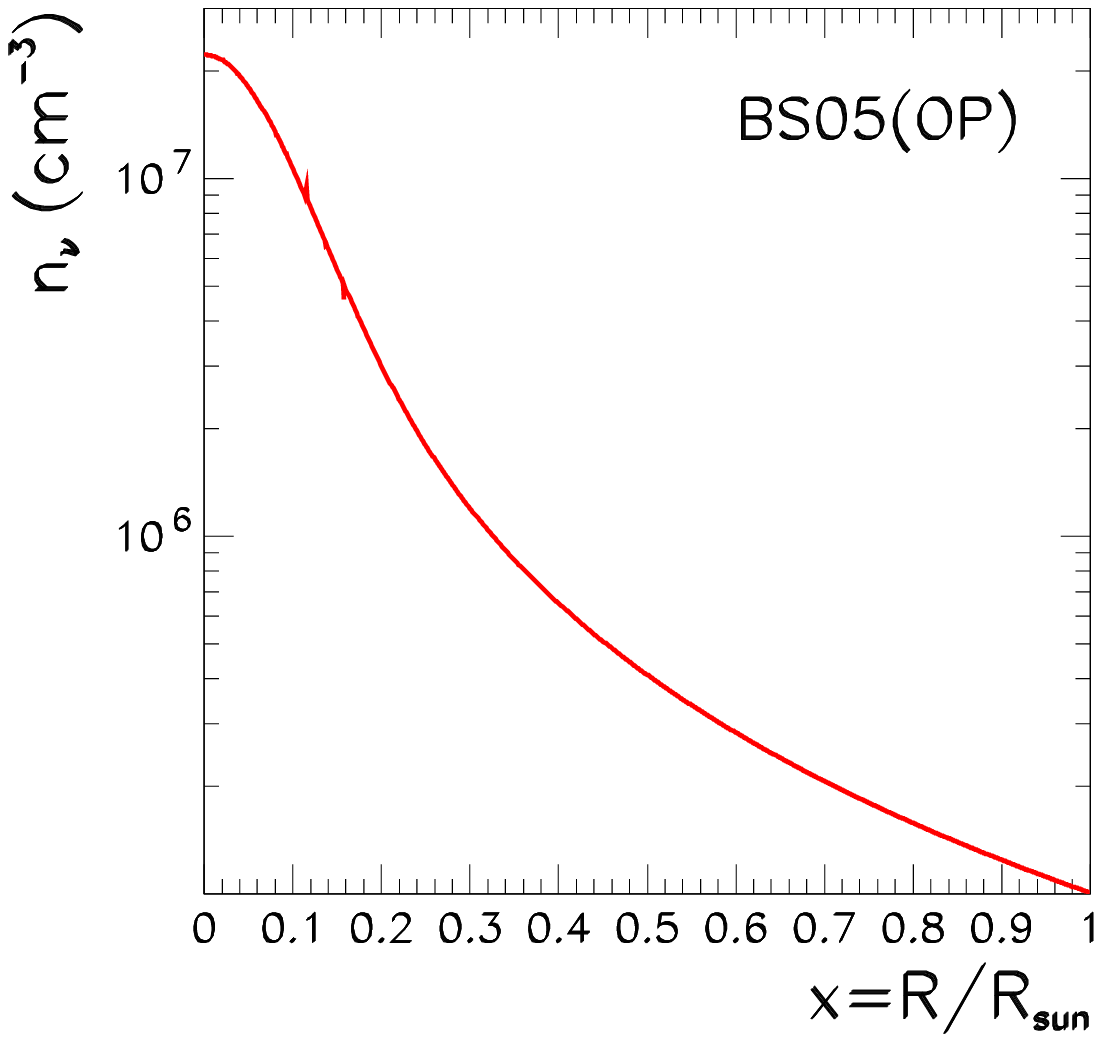}
\includegraphics[width=3.2in]{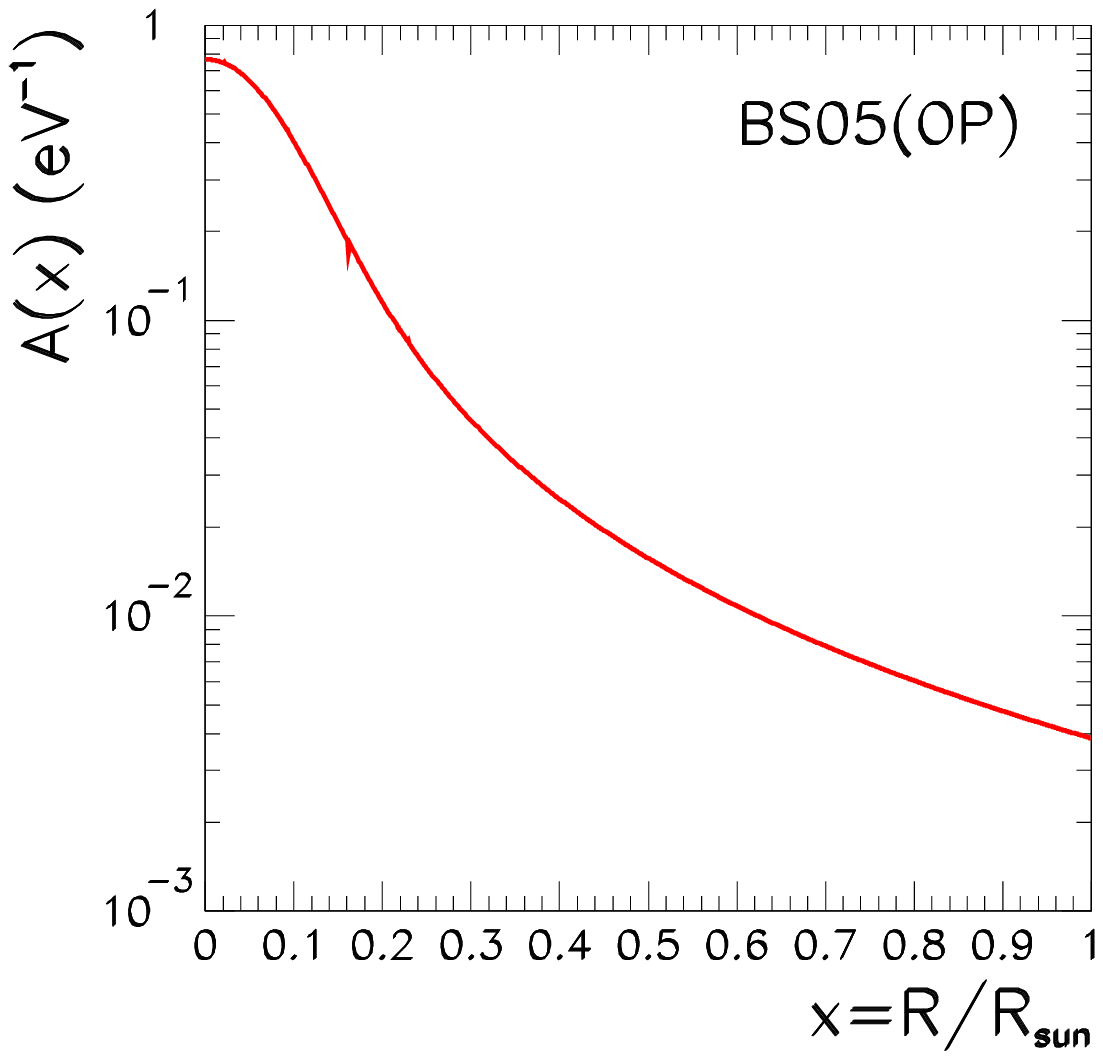}
\caption{Density of relativistic neutrinos in the Sun 
and the corresponding $A$ factor as a function of the 
distance from the center of the Sun.}  
\label{fig:nudens}
\end{figure}

Altogether we get the density of relativistic
neutrinos in the Sun shown in Fig.~\ref{fig:nudens}. 
As seen in the figure the neutrino density is maximum at the center of
the Sun where it reaches $2.2\times 10^7$/cm$^3$.
It decreases by over two orders of magnitude at 
the edge of the Sun.

Correspondingly we find their average inverse energy parameter normalized
to the CMB neutrino density (\ref{eq:a})
\begin{equation}
A(x)~=~ 0.00186~{\rm eV^{-1}}\, \frac{1} {x} ~\sum ~f_i~ 
\int d r ~ 4\pi 
 r ~\log \frac{x+ r}{|x-r|}~  p_i( r) \; ,
\label{eq:asun}
\end{equation}
where we have used 
\begin{equation}
\int dE \frac{1}{E}  \frac{d\Phi_{pp}}{dE}(E)= 
2.7 \times 10^{5} {\rm cm^{-2} s^{-1} eV^{-1}}\, , 
\end{equation}
and $f_i= \frac{\int dE \frac{1}{E}
~\frac{d \Phi_{\nu,i}}{dE}}{\int dE \frac{1}{E}
~\frac{d \Phi_{\nu,pp}}{dE}}=2.3\times 10^{-2},2\times 10^{-3},
 1\times 10^{-3},3.6\times 10^{-4},2.7\times 10^{-5}$, and $
4\times 10^{-6}$ give the small relative contribution from 
the $^7$Be, $N$, $O$, $pep$, $F$, and $^8$B fluxes.  
In deriving Eq.(\ref{eq:asun}) we have neglected the neutrino mass 
with respect to its characteristic energy in the Sun.

In Fig.~\ref{fig:nudens} we plot the factor $A(x)$ in    
Eq.(\ref{eq:asun}). As seen in the figure 
$A(x)\sim {\cal O}(1)$ eV$^{-1}$ in the region of maximum density,
as expected, since $A \sim (n^{\rm sun}/n^{C\nu B} )(1/\langle E_\nu\rangle)$
with  $\langle E_\nu\rangle\sim 0.1$~MeV being the characteristic
$pp$ neutrino energy.  The size of $A$ is what makes the effect so
relevant for solar neutrinos. Let us comment that Eq.(\ref{eq:asun})
is obtained under the approximation
that the energy spectrum of the neutrinos is independent
of the production point. This is a very good approximation since
the temperature inside the production region is known to vary
only within a factor $\sim 3$ ($T\sim$ 5--15 $10^{6} K$) 
which corresponds to energy variations 
of the order of keV. An extreme upper bound to the expected corrections
due to departures from this approximation can be obtained from 
the results of Ref.~\cite{jb91}. In that work the shapes of the 
different neutrino spectra in the solar interior and in the 
laboratory were compared and the corrections found were of the order
${\cal O}(10^{-5})$ for beta decay neutrino spectra and at most 1\% 
for the pp neutrino spectrum.

Solving Eq.~(\ref{eq:minimization}) with the $A(x)$ term above 
one finds the effective value 
of the  neutrino mass as a function of the solar neutrino density,
while the vacuum neutrino mass $m_\nu^0$ can be found from
the corresponding condition outside of any non-relic neutrino 
background
\begin{equation}
\frac{\partial V_{tot}(m^0_\nu)}{\partial m_\nu}\rfloor_{m_\nu^0} = 0
\quad \Rightarrow 
\quad  V'_0(m^0_\nu) +n^{C\nu B}=0 \, .
\label{eq:minimization0}
\end{equation}
It is clear from Eqs.~(\ref{eq:minimization}) and (\ref{eq:minimization0})
that the precise shift induced in the neutrino mass by the presence of
an additional neutrino density depends on the exact form of the 
scalar potential $V_0(m_\nu)$. In general one can parametrize the 
scalar potential as
\begin{equation}
V_{0}(m_\nu) = \Lambda^4 \ f\left(\frac{m_\nu}{\mu} \right)\, , 
\end{equation}
factoring out an overall scale $\Lambda^4$ 
which would set the scale of the
cosmological constant in a
standard scenario and a function $f$ which
depends on the dimensionless ratio $m_\nu/\mu$, where $\mu$ is an
accessory mass scale which will have no particular role for our discussion.
 
The observation that  the equation of state for the dark energy, 
\begin{displaymath}
\omega +1 = -\frac{m^0_\nu \ V'_{0}(m^0_\nu)}{V_{tot}(m^0_\nu)}\, ,
\end{displaymath}
must  have $\omega \approx -1$  
(e.g. $-1.21 <\omega < -0.88$ at 68\% c.l. 
combining the cosmological data sets~\cite{omega})
implies that the scalar potential must be fairly flat
\begin{equation}
\frac{dV_{0}(m_\nu)}{dm_\nu} \ll 1 \, .
\label{eq:flatness}
\end{equation}
Furthermore Eq.~(\ref{eq:minimization0}) implies 
\begin{equation}
\frac{dV_{0}(m_\nu)}{dm_\nu} <0\, ,
\label{eq:mono}
\end{equation}
this is,  the potential must be a monotonically decreasing
function of $m_\nu$. 

Given the requirements (\ref{eq:flatness}) and (\ref{eq:mono}) 
three suitable paradigmatic 
forms of the function $f(m_\nu/\mu)$ have been proposed
\cite{dark1, cosmo consequences}.\\
(i) A {\it logarithmic} form
\begin{equation} f
\left(\frac{m_\nu}{\mu}\right) =
\log\left(\frac{\mu}{m_\nu} \right).
\end{equation}
In this case  from Eqs.~(\ref{eq:minimization}) and (\ref{eq:minimization0})
one gets the equation for the neutrino mass shift
\begin{equation}
m_\nu-m^0_\nu=-A\ m^2_\nu 
\label{eq:logeq}
\end{equation}
whose solution in the  limit of small  $A$ is
\begin{equation}
m_\nu = m^0_\nu -A (m^0_\nu)^2 + \ldots 
\label{eq:logsolution}
\end{equation}
Eq.~(\ref{eq:logeq}) shows explicitly that  the relative shift
in the neutrino mass due to the additional neutrino background 
$(m_\nu-m^0_\nu)/m_\nu$ grows in magnitude with the neutrino mass scale.
%

\noindent (ii) A {\it power law} with a small fractionally power 
\begin{equation}
f\left(\frac{m_\nu}{\mu}\right) =
\left(\frac{m_\nu}{\mu}\right)^{-\alpha} \qquad  (\alpha >0) .
\end{equation}
The condition $\omega
\approx -1$ implies $\alpha \ll 1 $ and one gets
\begin{equation}
m_\nu-(m^0_\nu)^{\alpha+1}m_\nu^{-\alpha}=-A\ m^2_\nu  \, ,
\end{equation} 
which for  $\alpha \ll 1$ is the same as Eq.(\ref{eq:logeq}).
%

\noindent(iii) An {\it inverse exponential}
\begin{equation} f\left(\frac{m_\nu}{\mu}\right) =
e^{\frac{\mu}{m_\nu}}\, , 
\end{equation} 
implies 
\begin{equation} 
m_\nu -m^0_\nu \left( \frac{m^0_\nu}{m_\nu} \exp \left[-\frac{\omega +
1}{\omega}\left(\frac{m^0_\nu}{m_\nu}-1\right)\right] \right) =-A\
m^2_\nu \, ,
\end{equation}
which in the limit $\omega \rightarrow -1$ gives a cubic
equation in $m_\nu$ 
\begin{equation} m_\nu^2 - (m^0_\nu)^2 = -A\ m_\nu^3 \, , 
\end{equation} 
whose exact solution for small $A$ is 
\begin{equation} m_\nu 
=m^0_\nu -\frac{A}{2} (m^0_\nu)^2 + \ldots \ , 
\end{equation}
to be compared with Eq.(\ref{eq:logsolution}).

In summary within choices of the scalar potential which verify 
the conditions of flatness and monotony the relative shift in the neutrino
mass value due to the solar neutrino density background grows
with the neutrino mass while  
the exact value of the shift is only moderately model dependent. 

\section{Mass Varying Neutrino Oscillations in the Sun}
\label{sec:osc}
%
The discussion in the previous section applies to one 
neutrino species. In order to determine the effect of the
scenario on the solar neutrino oscillations we need to 
extend it to two or more neutrinos. This rises the issue of 
how many neutrino states do acquire a contribution to their mass
via the coupling to the acceleron field. In principle with 
one acceleron field, only one combination of the different 
$m_{\nu_i}\equiv m_i$ has to be taken to be the dynamical field 
for the purpose of analyzing the minimal energy density.

Notwithstanding, in the following discussion we are going to 
assume that all neutrinos acquire a contribution to their mass via the
couplings to the dark sector and that such contributions 
are independent\footnote{A trivial realization of such scenario 
is to introduce several stable acceleron fields which couple independently
to the different neutrino states.}. 

In this case we can simply write the effective Lagrangian for
the neutrinos as
\begin{equation} 
\mathcal{L}= \sum_i m_i \bar\nu^c_{i}
\nu_{i} + \sum_i\left[ m_i \ n^{C\nu B}_i + V_{\nu_i, \rm medium}
+ V_0(m_i)\right]\, , 
\end{equation}
and the condition of minimum of the effective potential 
implies that it has to be  verified that  
\begin{equation}
\frac{dV_0(m_i)}{dm_i}+ 
n^{C\nu B}_i + m_i\int 
\frac{d^3 k}{(2\pi)^3} \frac{1}{\sqrt{k^2 + m_i^2}} 
f_{\rm Sun,i}(k)=0\, ,
\label{eq:min3}
\end{equation}
for each $m_i$ independently. Under this assumption, the coupling
to the dark sector leads to a shift of the neutrino masses but does
not alter the leptonic flavour structure which is determined either
by other non-dark contributions to the neutrino mass or from 
the charged lepton sector of the theory. We will go back to this point 
after presenting the results.

For the sake of concreteness we will present our results on  
solar neutrinos oscillations for the case of a logarithmic 
potential $V_0(m_i)=\Lambda^4\log\left(\mu/m_i\right)$. 
In this case Eqs.(\ref{eq:min3}) lead to  
three (one for each neutrino) independent equations for the mass shifts
\begin{equation}  
(m_i-m^0_i)=-m_i^2~ A_i\, , 
\end{equation}
where 
\begin{equation}
A_i=\frac{1} {n^{C\nu B}_i} 
\int \frac{d^3 k}{(2\pi)^3} \frac{1}{\sqrt{k^2 + m_i^2}} \,
f_{\rm Sun,i}(k) \, .
\end{equation}
So even in this case of no leptonic mixing from the scalar potential, 
there is a generation dependence of the $A$ factor from the flavour
dependence of the background neutrino density.

We assume that all massive neutrinos have the
same contribution to the cosmic density,  
$n^{C\nu B}_i=112$ cm$^{-3}$ for all $i$. 
In this case the generation dependence comes from the fact
that in the Sun only  $\nu_e$'s are produced. Using the standard labeling of
the massive neutrino states and neglecting $\theta_{13}$ 
we find only the states $\nu_1$ and $\nu_2$ 
have their masses modified by the presence of the solar 
neutrino background as given in Eq.~(\ref{eq:logeq}) with  
\begin{equation}
\begin{array}{l}
n_1(x)~=~\cos^2\theta^V_{12}~n_{\nu_e}(x)~\Rightarrow~ A_1(x)~=
~\cos^2\theta^V_{12}~ A(x)\, ,\\
n_2(x)~=~\sin^2\theta^V_{12}~n_{\nu_e}(x)~\Rightarrow~ 
A_2(x)~=~\sin^2\theta^V_{12}~ A(x)\, ,
\end{array}
\end{equation} 
where $\theta^V_{12}$ is the vacuum mixing angle and $A(x)$ is
given in Eq.(\ref{eq:asun}). 

\begin{figure}[ht]
\includegraphics[width=3.2in]{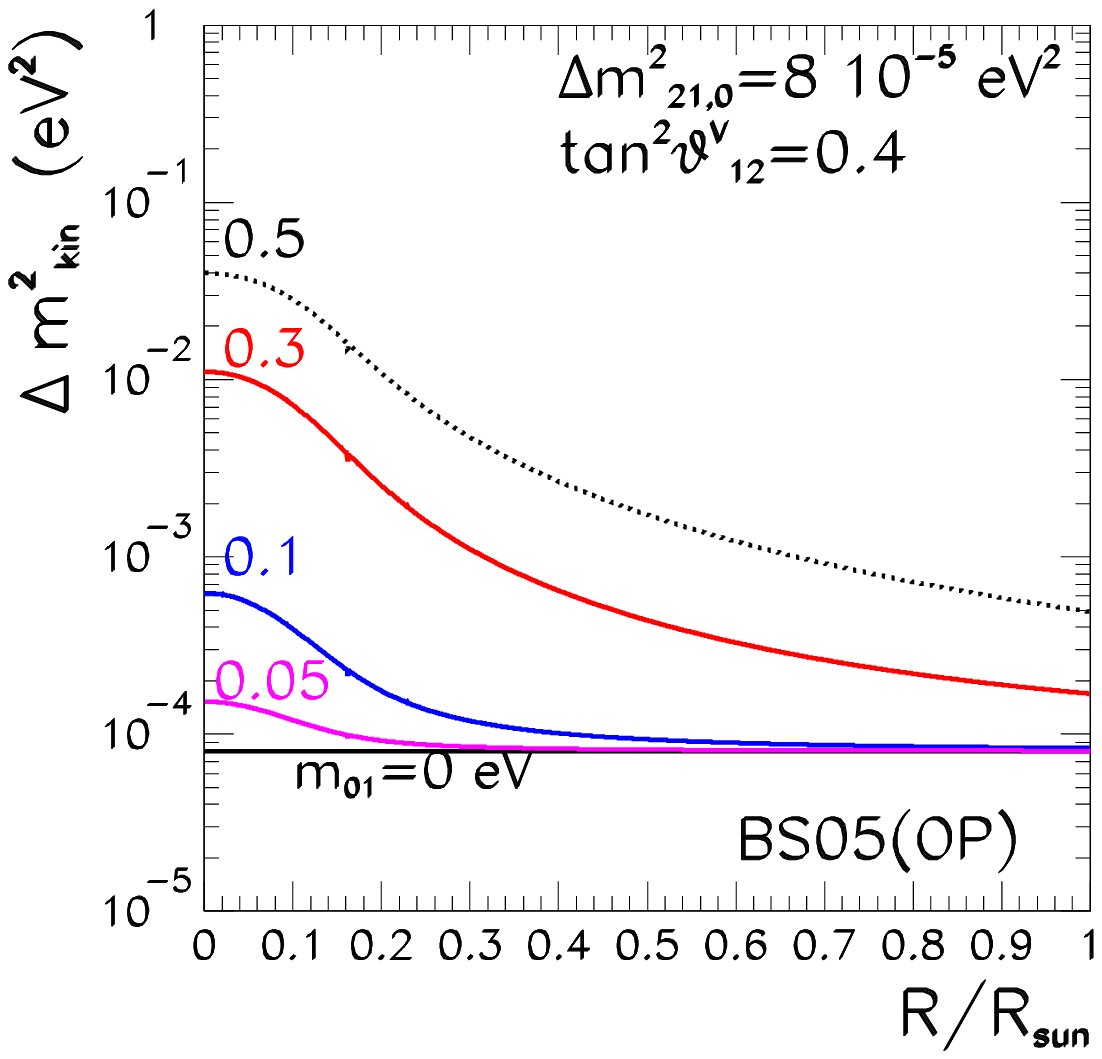}
\includegraphics[width=3.2in]{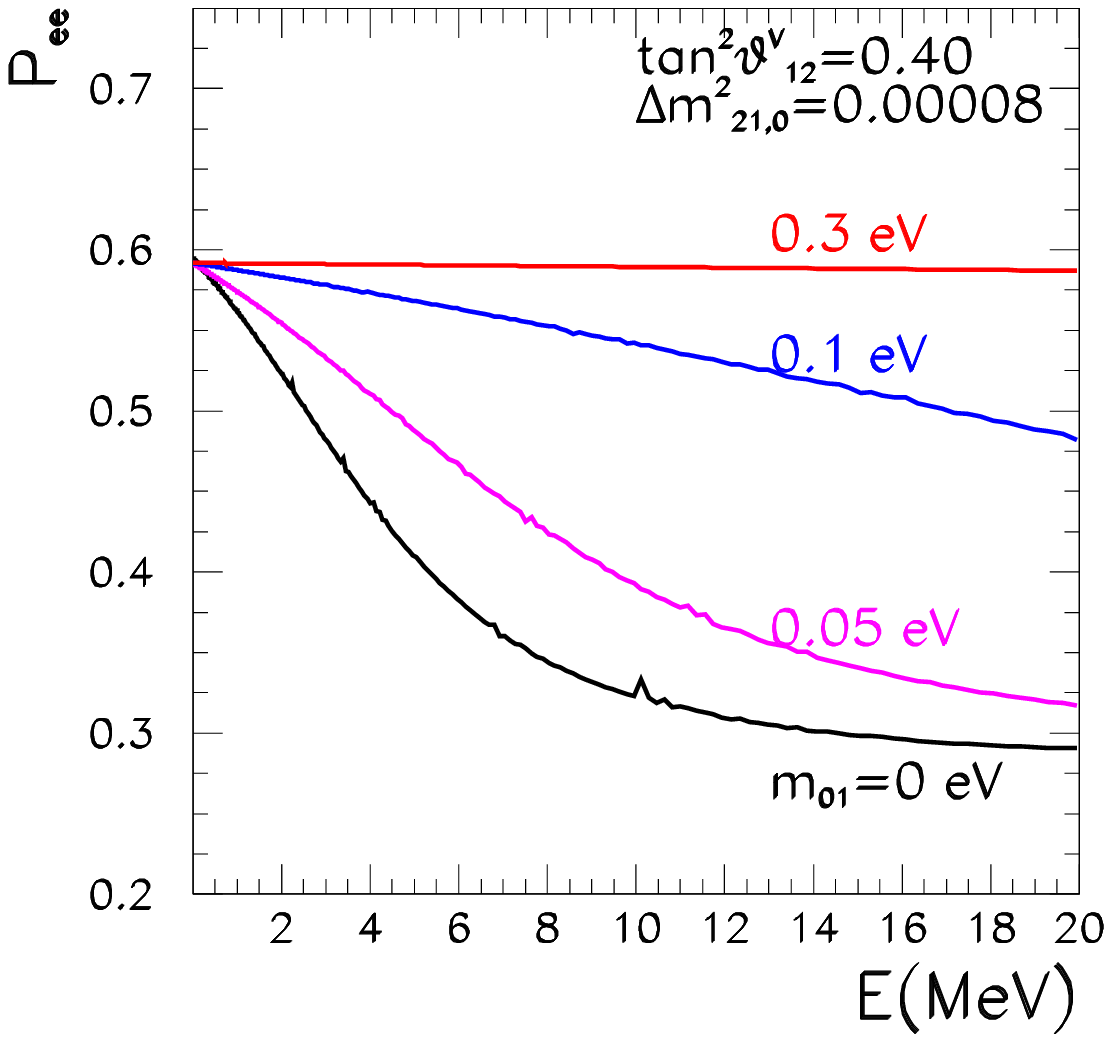}
\caption{(Left) Effective mass difference in the Sun.
(Right) Survival probability of solar $\nu_e$'s as a function
of the neutrino energy. This survival probability has been 
obtained for neutrinos produced around $x=0.05$ as it is characteristic
of $^8$B neutrinos.}
\label{fig:prob}
\end{figure}

Altogether this implies that the effective ``kinetic'' (we label it
kinetic to make it explicit that it does not contain the MSW potential) 
mass difference in the Sun is
\begin{equation}
\Delta m^2_{\rm kin}(x)= m^2_2(x)-m_1^2(x)\simeq 
\Delta m^2_{21,0}[1- 3 A_2(x) m_{01}] + 2 [A_1(x)-A_2(x)] m_{01}^3 +\dots
\label{eq:deltam}
\end{equation}
where, for clarity, we have given the explicit expression when 
expanded in powers of $A(x)$ and the neutrino mass scale $m_{01}$.
$\Delta m^2_{21,0}=m_{02}^2-m_{01}^2$ and
$\theta^V_{12}$ are ``vacuum'' mass difference and mixing angle. 
These are the parameters measured with reactor antineutrinos at KamLAND
\footnote{We estimate an $A$ factor from the background density of 
reactor antineutrino and geoneutrinos to be of the order of 
${\cal O}(10^{-11}{\rm eV}^{-1})$. This includes geoneutrinos from 
radioactive elements yielding (anti)neutrinos that are under 
the threshold of running experiments but actually give the dominant 
contribution to this $A$ factor.}. 
In Fig.~\ref{fig:prob} we plot the effective $\Delta m^2_{\rm kin}(x)$ 
as a function of the distance from the center of the Sun for different
values of the neutrino mass scale $m_{01}$. In this figure, and it what
follows the results with  $m_{01}=0$  are obtained by zeroing the
dark-energy contributions so $\Delta m^2_{\rm kin}(x)=\Delta m^2_{21,0}$. 
Strictly speaking our derivation of $\Delta m^2_{\rm kin}(x)$ assumes
that all CMB neutrinos are non-relativistic in the present epoch,
an assumption which does not hold for the lightest neutrino if $m_{01}=0$.
But as long as the behaviour is continuous, the contribution 
will be negligible small for this case.
 
From Eq.(\ref{eq:deltam}) we read that, as long as the different massive
neutrinos have different projections over $\nu_e$ ($A_1\neq A_2$),  
$\Delta m^2_{\rm kin}(x)$ receives a contribution from the solar 
neutrino background which rapidly grows with the neutrino mass scale 
$m_{01}$. For the particular scenario that we are studying 
$A_1(x)-A_2(x)=\cos 2\theta^V_{12}\, A(x)>0$ so
the effective kinetic mass splitting is positive and 
larger than the vacuum one in the resonant side for neutrinos .

Next we evaluate the corresponding survival probability for solar 
MaVaNs  by solving the evolution  equations 
\begin{equation}  
i{\displaystyle\frac{d}{dx}}  
\left( \begin{array}{c} \nu_e \\ \nu_\mu
 \end{array} \right) 
=  
\left[ \frac{1}{2E}\, U 
\left(\begin{array}{cc} 0 & 0 \\
0 & \Delta m_{\rm kin}^2 (x)
\end{array} \right) 
U^\dagger +
\left( \begin{array}{cc} 
V(x) & 0 \\
0 & 0 
\end{array} \right) \right]
\left( \begin{array}{c} 
\nu_e \\ \nu_\mu 
\end{array} \right).
\label{eq:evol} 
\end{equation}
where $V(x)=\sqrt{2} \,G_F N_e(x)$ is the MSW potential~\cite{MSW}.
We need not include MSW-like modifications 
induced by an effective one loop coupling neutrino-electron 
mediated by the acceleron, because these can be seen to be 
negligibly small under mild assumptions, as discussed 
in~\cite{dark1,dark2}. 
$U$ is the mixing matrix of angle $\theta^V_{12}$. We solve
this equation by numerical integration  along the neutrino trajectory.
However in most of the parameter space the evolution of the neutrino
system is adiabatic and the survival probability is very well reproduced
by the standard formula
\begin{equation}
P_{ee}=\frac{1}{2}+\frac{1}{2} \cos2\tilde\theta_{12,0}\cos2\theta^V_{12}\, ,
\end{equation}
where $\tilde\theta_{12,0}$ is the effective mixing angle at the neutrino
production point $x_0$. It includes both the effect of the point dependent
kinetic mass splitting  as well as the effect of the MSW potential. 
\begin{equation}
\cos2\tilde\theta_{12,0}=
\frac{\Delta m^2_{\rm kin}(x_0)\cos 2\theta^V_{12}-A_{\rm MSW}(x_0)}
{\sqrt{(\Delta m^2_{\rm kin}(x_0)\cos 2\theta^V_{12}-A_{\rm MSW}(x_0))^2
+(\Delta m^2_{\rm kin}(x_0)\sin 2\theta^V_{12})^2}}
\label{eq:mixing}
\end{equation}
where $A_{\rm MSW}(x_0)=2 E V(x_0)$.

We plot in Fig.~\ref{fig:prob} the survival probability as a function
of the neutrino energy for $\Delta m^2_{21,0}=8\times 10^{-5}$ eV$^2$ and
$\tan^2\theta^V_{12}=0.4$ and different values of the neutrino mass
scale $m_{01}$.  As can be seen in the figure, due to the different
contributions of the solar neutrino background to the two mass
eigenstates, the energy dependence of the survival probability is 
rapidly damped even for mildly degenerated neutrinos. 
As a consequence, in these cases,  it is not  possible to  simultaneously 
accommodate the observed event rates in solar neutrino 
experiments~\cite{chlorine,sagegno,gallex,sk,sno}
and in KamLAND ~\cite{kamland} as we quantify next.

\section{Constraints From Solar Neutrino Observables}
\begin{figure}[ht]
\includegraphics[width=5.5in]{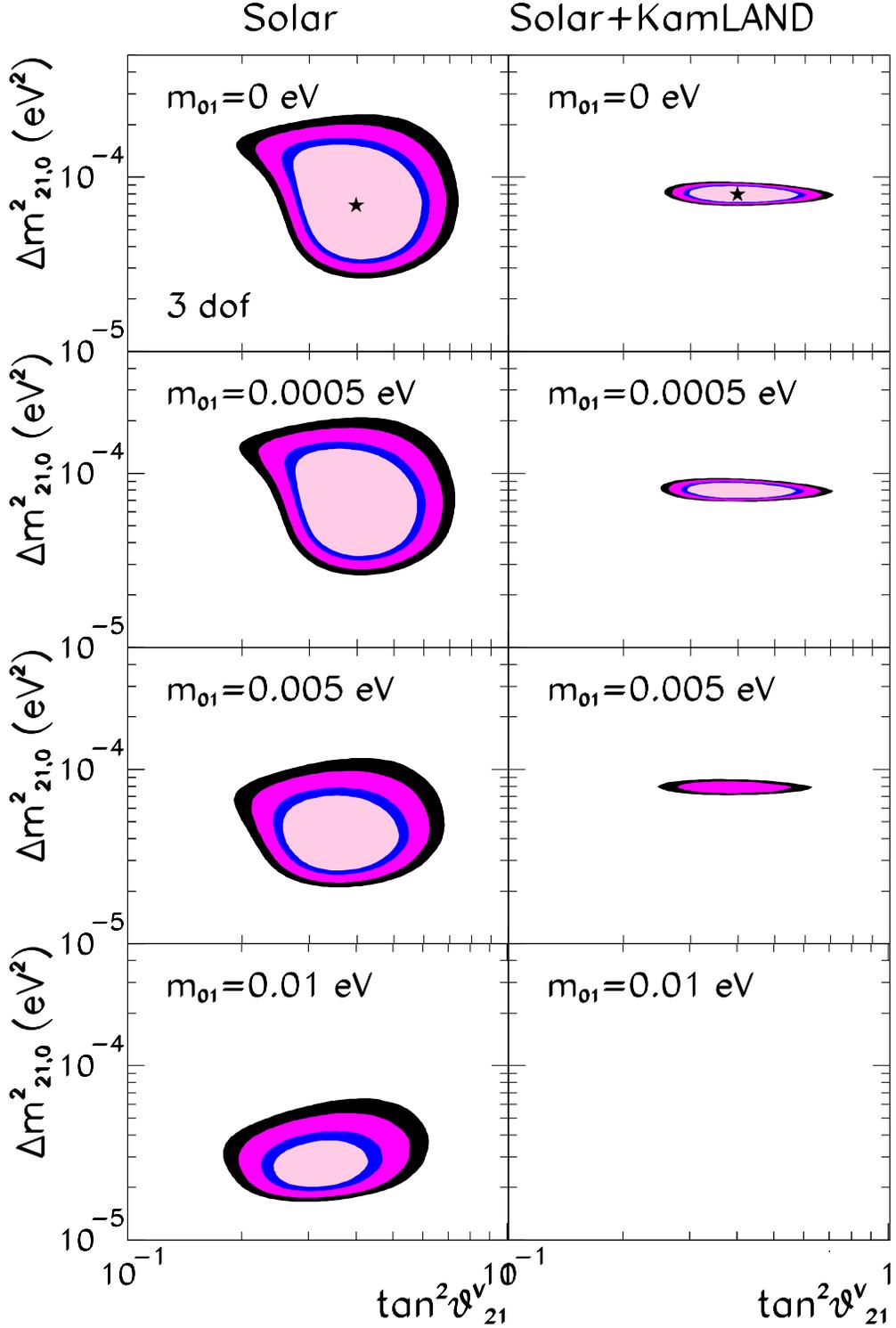}
\caption{Allowed regions from the global analysis of solar 
and solar plus KamLAND data in the 
$(\Delta m^2_{21,0},\tan^2\theta_{12}^V, m_{01})$ parameter space, shown
for 4 sections at fixed values of $m_{01}$. 
The different contours corresponds to 90\%, 95\%, 99\%, and 3$\sigma$ CL
for 3dof. The global minima are marked with a star.} 
\label{fig:regions}
\end{figure}\label{sec:analysis}

We present in this section the results of the global analysis of
solar and KamLAND data in the framework of MaVaNs 
for the specific 
realization discussed in the previous section. 

Details of our solar neutrino analyses have been described in previous
papers~\cite{oursolar,ourkland}. The solar neutrino data we
use includes the Gallium~\cite{sagegno,gallex} (averaged to 1 data point)
and Chlorine~\cite{chlorine} (1 data point)
radiochemical rates, the Super-Kamiokande~\cite{sk} zenith spectrum (44 bins),
and SNO data previously reported for phase 1 and phase 2.  The SNO
data used consists of the total day-night spectrum
measured in the pure D$_2$O phase (34 data points), plus the total
charged current (CC, 1 data point), electron scattering (ES, 1 data
point), and neutral current (NC, 1 data point) rates measured in the
salt phase~\cite{sno}. The main difference with respect to previous
analysis is that we use the solar fluxes from 
Bahcall, Serenelli and Basu 2005~\cite{BS05} but we still 
allow the normalization  of the $^8$B flux to be a free parameter 
to be fitted to the data.

The analysis of solar neutrino depends then of 4 parameters
$\Delta m^2_{21,0}, \tan^2\theta^V_{12}, m_{01}$, and $f_{\rm B}$
(the reduced flux $f_{\rm B}$, is defined as the $^8$B solar neutrino flux
divided by the corresponding value predicted by the BS05 standard
solar model).

We show in the left panels of Fig.~\ref{fig:regions} the result of the
global analysis of solar data in the form of the allowed
regions in the 3-dimensional parameter space of 
$\Delta m^2_{21,0}, \tan^2\theta^V_{12}, m_{01}$,
after marginalization over the  $f_{\rm B}$. 
The regions have been defined by the conditions 
$\Delta\chi^2_{\rm sol}(\Delta m^2_{21,0},\theta^V_{21},m_{01})\equiv
\chi^2_{\rm min,f_{\rm B}}(\Delta m^2_{21,0},\theta^V_{21},m_{01})
-\chi^2_{\rm min}\leq \Delta\chi^2 \mbox{(C.L., 3~d.o.f.)}$ , 
where $\Delta\chi^2(\mbox{C.L., 3~d.o.f.}) = 6.25$, $7.81$, $11.34$, and
$14.16$ for C.L.~= 90\%, 95\%, 99\% and 99.73\% ($3\sigma$)
respectively, and $\chi^2_{\rm min}$ 
is the global minimum which is
obtained for the totally hierarchical case $m_{01}=0$ eV and
$\Delta m^2_{21,0}=6.9\times 10^{-5}$ eV$^2$, 
$\tan^2\theta^V_{12}=0.4$ and $f_{\rm B}=0.92$.

In the figure we plot sections of the 3-dimensional allowed regions 
at fixed values of $m_{01}$. As seen in the figure, as $m_{01}$
increases the allowed region of the solar analysis shifts to
lower values of $\Delta m^2_{21,0}$ to compensate for  the increase of
$\Delta m^2_{\rm kin}$ and the fit to solar data worsens.
The worsening is driven by two main effects. First, 
the increase of the survival probability of $^8$B 
neutrinos makes more difficult to accommodate the observed 
CC/NC ratio (and CC/ES) at SNO. In principle the CC rate 
could be cured by the free $^8$B flux $f_{\rm B}$,  
but the NC constrains the allowed values of $f_{\rm B}$.
Second, shifting to lower values of $\Delta m^2_{21,0}$  
increases the expected day-night asymmetry. This eventually makes
the agreement with the data impossible for high enough values of
$m_{01}$ since the neutrino density in the Earth is too small
to induce any additional effect on the day-night asymmetry. 
Consequently, we find that the  3-dimensional region  at 3$\sigma$  
extends only to $m_{01}\leq 0.05$ eV.  
 
It is clear from these results, that the fit for large values of
$m_{01}$ will become worse after combination with the KamLAND data. 
In the present framework the analysis of KamLAND only depends on the 
``vacuum'' parameters $\Delta m^2_{21,0}$ and $\tan^2\theta^V_{21}$.
We include here the results of a likelihood analysis  
to the unbinned KamLAND data~\cite{klandhp}. Details of this
analysis will be presented elsewhere~\cite{carloskland}.

We show in the right panels of Fig.~\ref{fig:regions} the result of the
combined analysis of solar plus KamLAND. 
The global minimum is obtained for the totally hierarchical 
case $m_{01}=0$ eV and $\Delta m^2_{21,0}=7.9\times 10^{-5}$ eV$^2$, 
$\tan^2\theta^V_{12}=0.4$ and $f_{\rm B}=0.90$.
As seen in the figure as $m_{01}$ increases the allowed region becomes
smaller. As a matter of fact, due to the shift of the solar region to
lower values of $\Delta m^2_{21,0}$, the local best fit point of the 
combined analysis moves to the  LMA0 region~\cite{carlosnsi} 
for ``intermediate'' values of $m_{01}\sim {\cal O}(10^{-2})$ eV.  
In other words, for
those values, LMA0 becomes less disfavoured than in the hierarchical
case. For example for $m_{01}=0$ the LMA0 region lies at
$\Delta\chi^2=37.5$, which implies that it would be part of
the 3-dim allowed region at 5.5$\sigma$, while for $m_{01}=0.01$ eV the 
LMA0 region  lies at $\Delta\chi^2=15.3$ and it would be part of the 
3-dim region at  allowed at 3.15$\sigma$. 

The result of the previous discussion is that generically MaVaN's 
imply that the description of solar data worsens with the degree
of degeneracy of the neutrinos. In order to quantify this statement
in the present scenario, we define the ``degeneracy parameter'', 
$x_{\rm deg}\equiv\Delta m^2_{21,0}/m^2_{01}$, and study the dependence of 
$\chi^2$ on this parameter after marginalization over all others:
\begin{equation}
\Delta\chi^2_{\rm sol(glob)}(x_{\rm deg})=
{\rm min}\chi^2_{\rm sol(glo)}(\Delta m^2_{21,0},\theta_{12}^V,
m_{01},f_{\rm B}|x_{\rm deg}=\Delta m^2_{21,0}/m^2_{01})-
\chi^2_{\rm min, sol(glob)}\, .
\end{equation}
\begin{figure}[ht]
\includegraphics[width=4in]{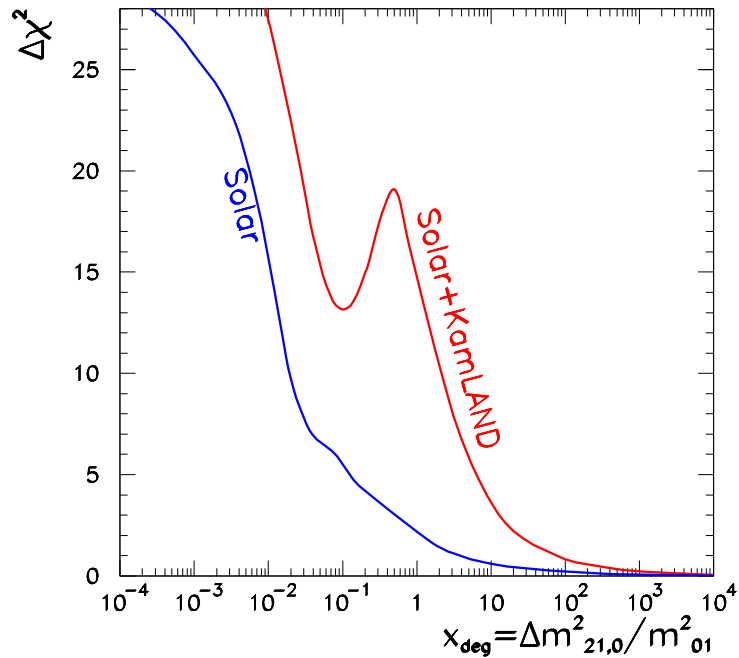}
\caption{Dependence of $\Delta\chi^2$ on the degeneracy parameter
$\Delta m^2_{21,0}/ m_{01}$ from the global analysis of 
analysis of solar  and solar plus KamLAND data after marginalization
in all other parameters.}  
\label{fig:deg}
\end{figure}

In Fig.~\ref{fig:deg} we plot  $\Delta\chi^2_{\rm sol(glob)}(x_{\rm deg})$.
Within the present bounds on the absolute neutrino 
mass~\cite{numass}, 
$2\times 10^{-5} \lesssim x_{\rm gen}<\infty$. 
As discussed above, we find that for the considered scenario of MaVaN's, 
the best fit occurs for hierarchical neutrinos $x_{\rm deg}=\infty$ 
while the fit becomes worse as the neutrinos become more degenerate.
As seen in the figure, the curve for the solar plus KamLAND analysis 
is not monotonic but presents a secondary minimum around 
$x_{\rm deg}=0.1$. This is due to
the migration of the local best fit point to the LMA0 region for  
values of $m_{01}\sim {\cal O}(10^{-2})$ eV. 
 
Quantitatively, we find  the lower bound at 3$\sigma$:
\begin{equation}
x_{\rm deg}> 2\times 10^{-2}\, (1) \, ,
\end{equation}
from the analysis of solar (solar plus KamLAND) data. 
In particular, this bound implies that in this scenario 
inverted mass ordering is disfavoured since in this
case  $m_{01}^2\simeq \Delta m^2_{\rm ATM} \gtrsim 10^{-3}$ eV$^2$ 
which implies $x_{\rm deg}\lesssim 0.1$. 

Finally we want to comment on the possible model-dependence of these
results.  As discussed in the previous sections there are two main 
sources of arbitrariness in our derivations: the choice of the functional
form of the scalar potential, and the assumption that all neutrinos
acquire an independent contribution to their mass via the couplings to
the dark sector with no generation mixing.

As shown in our discussion in Sec.~\ref{sec:mass} the choice of the
potential may affect the exact form of the equation relating
$\Delta m^2_{21,0}$ to $\Delta m^2_{kin}(x)$ but it will not alter
the fact that  $\Delta m^2_{kin}(x)$ grows with the neutrino mass
scale. In particular choosing a  power law potential 
with a small fractionally power $\alpha\ll 1$ yields the same results 
while the results for an inverse exponential potential 
are very similar but for a slightly higher value of $m_{01}$.

Concerning the assumption of no generation mixing from the dark sector
contribution to the neutrino mass, its effect can be understood as
follows.  In general, if the couplings to the dark sector are not
``mass--diagonal'' they will induce an additional source of rotation
between the flavour eigenstates and the effective mass
eigenstates. This would imply that the mixing angle in
Eq.(\ref{eq:deltam}) would not be $\theta_{12}^V$ but some
$\theta_{12}^{\rm kin}(x)$.  In general the qualitative features of
the results will still be valid although the quantitative 
bounds will obviously vary.  In particular, the bounds will
become tighter if the mixing could be such that the mass eigenstates 
were inverted ($\theta_{12}^{\rm kin}(x)> \pi/4$).  

A possible exception to this general argument would be the special
case in which the flavour structure of the potential is such that
$A_2(x)\simeq A_1(x)$ without a substantial modification of
$\theta_{12}$. In this case the $A_2(x)-A_1(x)$ term in
Eq.~(\ref{eq:deltam}) would be suppressed and the shift on $\Delta
m^2$ would be small even for $m_{01}\sim 2$ eV. This would imply $\Delta
m^2_{kin}(x)<\Delta m^2_{21,0}$, this effect being mostly relevant for
neutrinos which are produced nearer the center of the Sun.  As a
consequence the survival probability for $^7$B neutrinos can be
slightly lower and a slightly better fit to the data could be
achieved.  

Summarizing, in this work we have studied the phenomenological 
consequences of the dependence of MaVaNs  on the neutrino density 
in the Sun. We have evaluated the density profile of neutrinos in the 
Sun in the SSM and the expected size of the neutrino mass shift
induced for different forms of the scalar potential.  We find that 
generically these scenarios establish a connection between the 
effective mass splitting in the Sun and the absolute neutrino mass scale.
We have analyzed the quantitative consequences of this effect, by
performing a global analysis to solar and KamLAND data for a particular
realization of this mechanism. Our results show that the description of solar 
neutrino data worsens for large neutrino mass scale and an upper
bound on the absolute neutrino mass scale can be derived. 
Equivalently, we derive  a lower bound on the level of degeneracy
$\Delta m^2_{21,0}/m^2_{01}> 2\times 10^{-2}\, 
(1)$ from the analysis of solar (solar plus KamLAND) data. 
A straightforward consequence of this is that 
normal mass orderings are favoured over inverse mass orderings.    

These results, in combination with a positive determination of the 
absolute neutrino mass scale from independent means, can be used
as a test of these scenarios. Ultimately, these scenarios will
be tested by the precise determination of the energy dependence of 
the survival probability  of solar neutrinos, in particular
for low energies~\cite{lowe,oursolar}.

\acknowledgments
We thank John Bahcall, Maurizio Piai, Aldo Serenelli, and 
Neal Weiner for useful conversations. We are particularly indebted to
A. Serenelli for discussions on the detailed production point
distributions of neutrinos in the BS05 model.
The work of M.C. is supported in part by the
USA Department of Energy under contract DE-FG02-92ER-40704.
CPG acknowledges support from the Keck Foundation and 
NSF grant No.~PHY0354776.
MCG-G is supported by  National Science Foundation
grant PHY0098527 and by Spanish Grant No FPA-2004-00996.


\end{document}